\documentclass[journal]{IEEEtran}

\usepackage{graphicx}
\usepackage[T1]{fontenc}
\usepackage{amsmath}
\usepackage[cmintegrals]{newtxmath}
\usepackage{url,multirow,morefloats,floatflt,cancel,tfrupee}
\usepackage{dblfloatfix}
\usepackage{colortbl}
\usepackage{xcolor}
\usepackage{pifont}
\usepackage[nointegrals]{wasysym}
\begin{document}

\title{Single-Photon Infrared Imaging with a Silicon Camera Based on Long-Wavelength-Pumping Two-Photon Absorption}

\author{
	Jianan Fang, Yinqi Wang, E Wu, Ming Yan, Kun Huang, and Heping Zeng
	\thanks{Manuscript received $\times \times$; revised $\times \times$; accepted $\times \times$. Date of publication $\times \times$; date of current version $\times \times$. This work was supported in part by the Program for Professor of Special Appointment (Eastern Scholar) at Shanghai Institutions of Higher Learning, Science and Technology Innovation Program of Basic Science Foundation of Shanghai under Grant (18JC1412000), National Natural Science Foundation of China (11621404, 11727812), and Fundamental Research Funds for the Central Universities. \textit{(Corresponding author: Kun Huang.)}}
        \thanks{J, Fang, Y. Wang, E Wu, M. Yan, K. Huang, and H. Zeng are with State Key Laboratory of Precision Spectroscopy, East China Normal University, Shanghai 200062, China (e-mail: khuang@lps.ecnu.edu.cn).}
        \thanks{E Wu, M. Yan, K. Huang, and H. Zeng are with Chongqing Institute of East China Normal University, Chongqing 401121, China.}
        \thanks{H. Zeng is with Jinan Institute of Quantum Technology, Jinan, Shandong 250101, China.}
        \thanks{H. Zeng is with Shanghai Research Center for Quantum Sciences, Shanghai 201315, China.}
        \thanks{Color versions of one or more of the figures in this letter are available online at http://ieeexplore.ieee.org.}
        \thanks{Digital Object Identifier $\times \times \times$}
        }

\maketitle 

\begin{abstract}
We experimentally demonstrated an ultra-sensitive imaging system for telecom photons based on the non-degenerate two-photon absorption in a silicon-based electron multiplying charge-coupled device (EMCCD). The proposed long-wavelength-pumping scheme with mid-infrared pulsed excitation could not only effectively increase the two-photon absorption coefficient, but also significantly suppress the background noise caused by the harmonic absorption of the strong pumping field. In comparison to the photoelectric response via the degenerate two-photon absorption, the implemented configuration could offer over 30-folded enhancement of the photon-counting rate in the infrared imaging. The resulting detection sensitivity up to 1 photon/pixel/pulse was unprecedentedly approached, thus facilitating the single-photon operation. The elimination of the stringent phase matching as typically required in the optical parametric conversion has led to a high spatial resolution of 13 $\mu$m. Moreover, the on-chip nonlinearity of the optical imager would enable a broadband spectral window and an enlarged field of view. In combination with the 5-ps temporal resolution due to the coincident optical gating, the presented imaging system would find various promising applications, such as low-light fluorescence lifetime microscopy and photon counting time-of-flight 3D imaging.
\end{abstract}


\IEEEpeerreviewmaketitle

\section{Introduction}
Sensitive optical detection is highly demanded in numerous applications, ranging from fundamental quantum optics to applied industry fields \cite{Hadfield2009NP, Eisaman2011RSI, Zhang2015LSA}. Nowadays, there have been various instantiations proposed to improve the detection sensitivity, including Geiger-mode avalanche photodiodes (APD) \cite{Kang2009NP}, homodyne coherent detectors \cite{Runge2017JSTQ}, and enhanced photosensors by high-density quantum dots \cite{Umezawa2014JSTQ}, micro-size resonators \cite{Ren2019JSTQ} or nano-structured light trapping \cite{Zang2017NC}. Among existing single-photon detectors, silicon photodetectors attract great attention due to the superior performances like nearly-unitary quantum efficiency, extremely low dark count rate, sub-nanosecond time-stamping capability, and as high as gigahertz operation rate \cite{Kim2011RSI, Campbell2016JLT}. Moreover, the advanced silicon photonic integrated circuit (Si PIC) technology enables to implement single-photon detector arrays \cite{Bruschini2019LSA, Pavia2014JSTQ, Bronzi2014JSTQ} and electron multiplying CCDs (EMCCDs) \cite{Huang2012APL}, which are capable of ultra-sensitive imaging at the single-photon level.

Recognizing the attractive features of Si-based detectors, there has been a steady interest to extend the operation window from the visible band to the infrared region. However, the bandgap energy of the silicon is above the photon energy for the telecommunication window. The desirable infrared response for a silicon detector has been investigated by using the Ge on silicon, SiGe or defect implantation \cite{Kang2009NP}. Alternatively, the material nonlinearity of the detector itself could be exploited to generate free charge carriers by the two-photon \cite{Bristow2007APL, Reichert2016PRL} or multi-photon \cite{Hurlbut2007OL,Pearl2008APL, NevetOL2011} absorption. This strategy can avoid extra fabrication steps beyond the standard CMOS process, which is thus more cost-effective and easier to incorporate into the Si PICs. Moreover, the involved manipulation of charge carriers does not require the phase matching in contrast to the upconversion scheme based on the spectral transduction of the incident infrared field \cite{Tanzilli2005Nature, Huang2021PR, Barh2019AOP}, favoring a broadband spectral acceptance, a large field of view, and no need of optics alignment. 

\begin{figure*}[t!]
\centering
\includegraphics[width=0.8\textwidth]{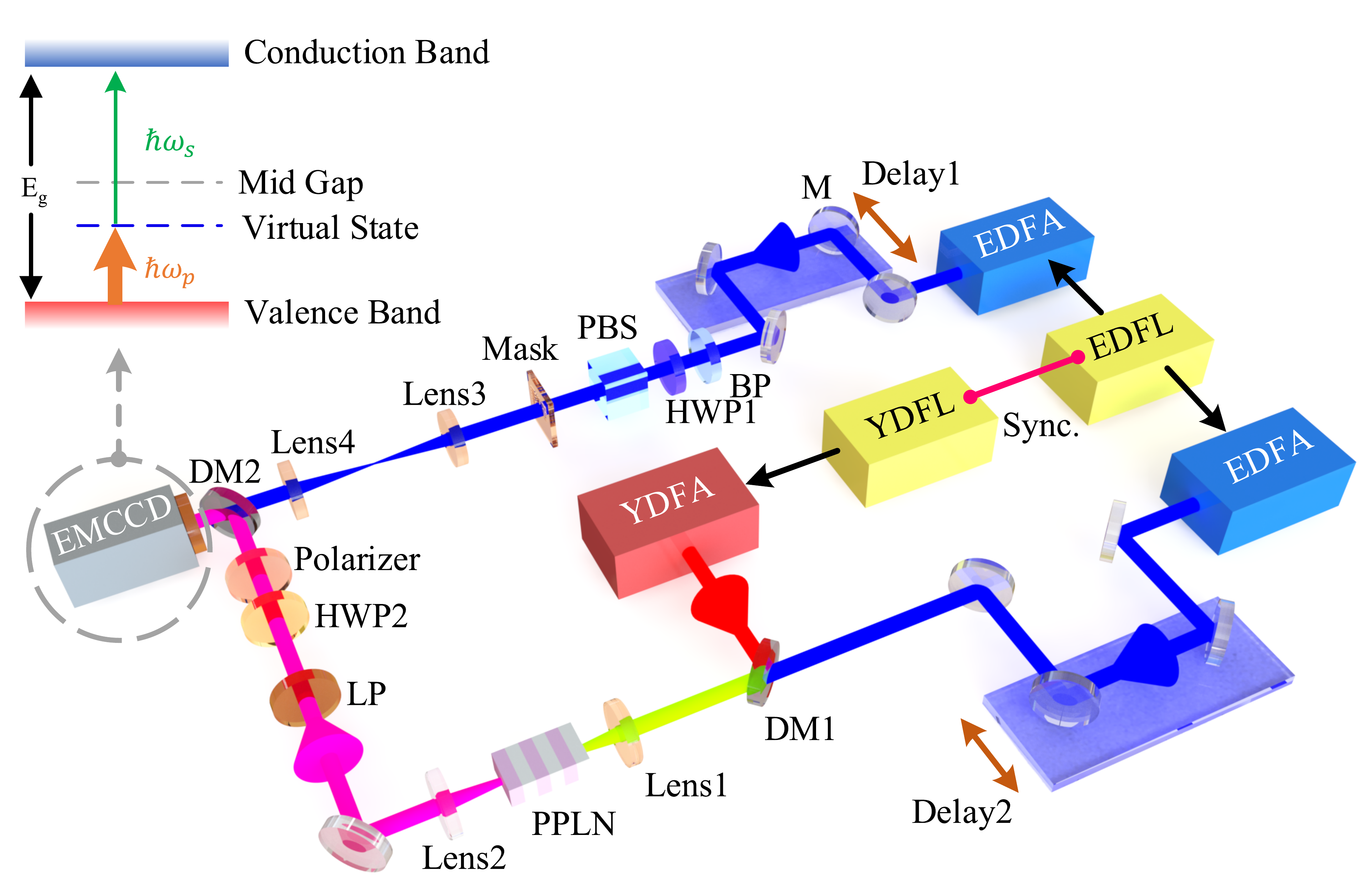}
\caption{Experimental setup for the 2PA-based infrared imaging in a silicon electron multiplying CCD (EMCCD) camera. The pump field at 3070 nm promotes an electron at the valence band to a virtual state, then the signal photon at 1550 nm completes the transition to a real conduction-band state. The pump photon energy of 0.4 eV is chosen to be lower than the semiconductor midgap about 0.56 eV as shown in the top-left energy diagram, which can eliminate the noise due to the degenerate 2PA in spite of the high pump intensity. The signal and pump pulsed sources originate from two Yb- and Er-doped fiber lasers (YDFL and EDFL). Particularly, the mid-infrared pump is prepared by the difference-frequency generation within a periodically poled lithium niobate crystal (PPLN). The illuminated mask and the camera constitute a 4f imaging system with two relay lenses. The signal and pump powers can be precisely controlled by combinating a polarizer and half-wave plate (HWP). EDFA and YDFA: Er- and Yb-doped fiber amplifier; BP and LP: band- and long-pass filter; PBS: polarization beam splitter; DM: dichroic mirror.}
\label{fig1}
\end{figure*}

In the context, there have been tremendous endeavors for optimizing the detection performance and extending subsequent applications, especially based on the two-photon absorption (2PA). For instance, the 2PA has long been used to implement the auto-correlators for ultrafast pulse characterization in the infrared region \cite{Boiko2017APL}. Recently, the pronounced enhancement of the nonlinear absorption coefficient has been observed in the extreme non-degenerate 2PA (ND-2PA) regime, which exhibited improvement factors of 100-1000 over the degenerate 2PA (D-2PA) \cite{Fishman2011NP,Cirloganu2011OE}. Consequently, the ND-2PA could provide a simple but effective way to realize sensitive infrared detection in wide-bandgap semiconductors, as been implemented based on a GaAs photomultiplier tube \cite{Boitier2009APL} and a Si APD \cite{Xu2019PTL}. Furthermore, two-dimensional infrared imaging was demonstrated by using a raster scanned GaN photodiode sensor \cite{Pattanaik2016OE} or a high-definition Si camera \cite{Knez2020LSA,Potma2021arXiv}. To go beyond the demonstrated sensitivity, a longer pump wavelength has been chosen to eliminate the severe background noises due to the pump harmonic excitation, whereas the parasitic process of three-photon absorption (3PA) could be negligible \cite{Fang2020PRA}. According to the seminal proposal \cite{Hayat2008PRB}, the long-wavelength-pumping scheme could approach to the single-photon sensitivity. Although such technique has been shown to facilitate highly-sensitive infrared detection with single-pixel detectors \cite{Fang2020PRA}, its advantages have not yet been translated to the direct imaging with high-performance Si-based cameras.

Here, we experimentally demonstrated the ultra-sensitive infrared imaging with a Si-based electron multiplying CCD based on the long-wavelength-pumping two-photon absorption. The non-degenerated operation with a pump photon energy below the silicon mid-gap could enhance the nonlinear absorption for the signal, and meanwhile eliminate the second-harmonic noise of the pump. The resultant improvement of the detection efficiency and the reduction of background noise enabled us to obtain an unprecedented imaging sensitivity with single-photon pulse energy on each pixel. Relative to previous works on the 2PA-based imaging, the achieved advance in the sensitivity has been made possible by combining the low-noise operation with the long-wavelength pumping and the photon-counting capability with the silicon EMCCD camera. Additionally, the nonlinear process of the two-photon resonance does not require the phase-matching condition, thus favoring the sensitive imaging with a high spatial resolution and a large field of view. Thanks to the gated operation of the imaging system, the longitudinal or depth resolution was determined by the gating pulse duration. In our experiment, the picosecond optical gate window resulted in a temporal resolution beyond the intrinsic timing jitter of the detector.
 
\section{Experimental setup}
Figure \ref{fig1} illustrates the experimental setup for the 2PA-based infrared imaging. The involved signal and the pump sources were prepared by using a temporally synchronized mode-locked fiber laser system, which consisted of an Yb-doped fiber laser (YDFL) at 1030 nm and an Er-doped fiber laser (EDFL) at 1550 nm. Details about the laser system can be referred to our previous work \cite{Zeng2019OL}. The all-polarization-maintaining synchronization system was implemented in an all-optical and passive configuration, thus eliminating the auxiliary feed-back complexity as required in the conventional active scheme. The presented system could thus provide a simple and compact solution to obtain the dual-color ultrafast pulses, which would be essential to realize the mid-infrared (MIR) light for the subsequent imaging based on the long-wavelength pumping. Specifically, the power of the synchronized pulses were firstly boosted by Yb- and Er-doped fiber amplifiers (YDFA and EDFA). Then the amplified dual-color pulses were steered into a periodically poled lithium niobate (PPLN) crystal to facilitate the preparation of MIR pulses at 3070 nm via the difference-frequency generation \cite{Huang2021HPLSE}. A translational stage (Delay2) was inserted in the optical path to optimize the conversion efficiency by tuning the temporal overlap. The resulting MIR light was spatially combined with the near-infrared (NIR) signal by a dichroic mirror before being sent into a silicon EMCCD camera (Andor, iXon Ultra 888). To suppress the ambient noise, an uncoated silicon window was placed in front of the camera. The infrared signal at 1550 nm was used to illuminate a transmission mask to form the object image, which was then mapped to the detection chip of the EMCCD through a 4f imaging system. The infrared image was made visible on the Si camera based on the ND-2PA. In the experiment, the pulse duration for the signal at 1550 nm and the pump at 3070 nm were inferred to be 1 ps and 5 ps according to the pulse correlation measurement. The longer pump envelop could completely enwrap the signal photons to improve the total detection efficiency. Additionally, the signal and pump powers could be precisely controlled by combing a polarizer and a half-wave plate.

\section{Results and discussion}
The underlying mechanism is depicted in the inset of Fig. \ref{fig1}, where the photocurrent in the Si semiconductor can be induced by the simultaneous absorption of the signal and pump fields. Specifically, the signal and pump photons have photon energies of 0.4 and 0.8 eV, and the combined photon energy of 1.2 eV exceeds the bandgap of 1.12 eV. In contrast to previous demonstrations for 2PA-based imaging \cite{Pattanaik2016OE, Knez2020LSA}, the pump wavelength was chosen to make sure that the pump photon energy was lower than the Si midgap of 0.56 eV. This unique setting could eliminate the background noise due to the harmonic excitation of the pump, thus leading to an improved sensitivity for the 2PA-based detector \cite{Fang2020PRA, Hayat2008PRB}. Additionally, the adopt of the non-degenerate operation could significantly enhance the nonlinear absorption \cite{Cirloganu2011OE}, which would favor the further increase of the signal-to-noise ratio.

In the semiclassical approach, the total photon count for all the pixels on the CCD camera is given by \cite{Boitier2009APL, Xu2019PTL}
\begin{equation}
N_\text{total} \approx N_\text{ND-2PA} + N_\text{D-2PA}  + N_\text{D-3PA}  \ , 
\label{eq1}
\end{equation}
where the $N_\text{ND-2PA}$,  $N_\text{D-2PA}$, and $N_\text{D-3PA}$ denote the non-degenerate response, two-photon absorption of the signal, and three-photon absorption of the pump. In the long-wavelength-pumping scheme, the last term is orders of magnitudes weaker than the two-photon process, and could thus be negligible \cite{Wang2013OE, Benis2020Optica}. In the presence of weak signal, the main contribution of the imaging signal is ascribed to the term $N_\text{ND-2PA}$ as the product of the two involved beams:
\begin{equation}
\begin{split}
N_\text{ND-2PA} & \propto \iint  I_p(x,y) \times I_s(x,y) dx dy   \\
& \propto I_p \times I_s \iint  e^{-2(x^2+y^2)/w^2} dx dy \\
& = \beta \times P_p \times P_s  \ ,
\end{split}
\label{eq2}
\end{equation}
where the powers of the pump and signal are given by $P_{p,s} \propto \iint I_{p,s} e^{-2(x^2+y^2)/w_{s,p}^2}dx dy $, the size of the ND-2PA beam is given by $w = w_s w_p/\sqrt{w_s^2 + w_p^2} $, and $\beta$ denotes the detection efficiency. It can be seen that the ND-2PA signal could be amplified by using a strong pump field as the local oscillator, which allows to detect extremely weak signals.

\begin{figure}[b!]
\centering
\includegraphics[width=0.95\columnwidth]{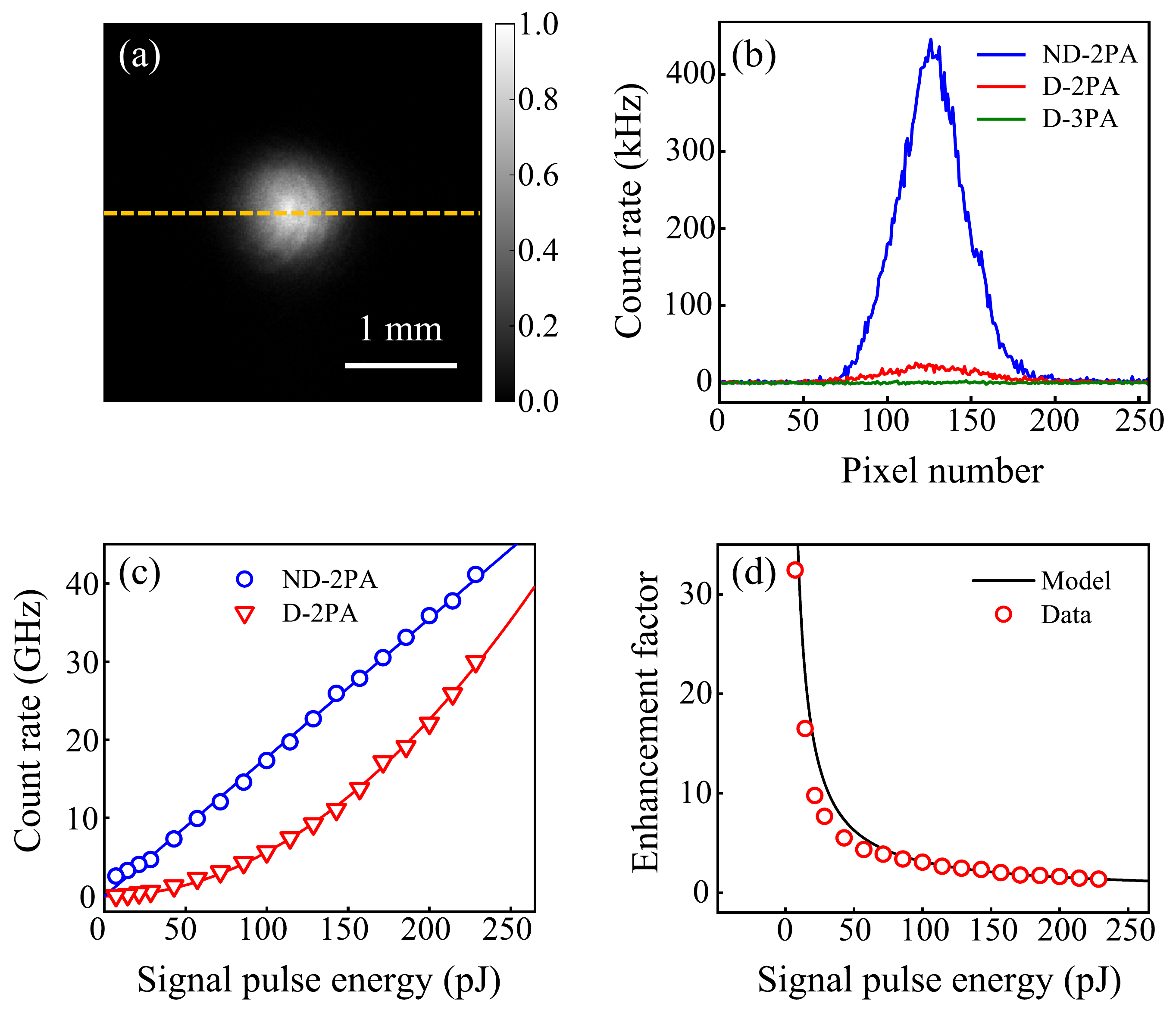}
\caption{(a) Infrared signal beam captured by the mid-infrared pumped silicon camera. The signal and pump pulse energy is about 7.1 pJ and 3.7 nJ, respectively. (b) Photon-counting rates for the pixels along the middle section at various illumination settings. A pronounced enhancement for the detection efficiency can be obtained by the ND-2PA, while D-3PA background noise is significantly suppressed. (c) Count rate by summing up all the registered photons in the image as a function of the signal pulse energy in the degenerate (D-2PA) and non-degenerate (ND-2PA) regimes. (d) Enhancement factor versus the signal pulse energy by comparing the count rates between the ND-2PA and D-2PA.} 
\label{fig2}
\end{figure}

In comparison, the D-2PA response of the signal beam can be written as
\begin{equation}
N_\text{D-2PA}  =  \gamma \times P_s^2  \ ,
\label{eq3}
\end{equation}
where $\gamma$ denotes the detection efficiency in the presence of only the signal beam. According to Eqs. (\ref{eq2}) and (\ref{eq3}), the enhancement factor for the two detection scheme can be defined as
\begin{equation}
G =  N_\text{ND-2PA}/N_\text{D-2PA} \propto 1/P_s  \ .
\label{eq4}
\end{equation}
Indeed, the ND-2PA scheme is particularly suitable for detecting weak signals. In this case, a more pronounced enhancement could be obtained for a lower signal intensity because of the quadratic decrease of the D-2PA contribution.

In the following, we characterized the imaging performance. Figure \ref{fig2}(a) presents the captured image with a signal and pump pulse energy of 7.1 pJ and 3.7 nJ, respectively. The intensity distribution along the middle section was given in Fig. \ref{fig2}(b). In comparison, the D-2PA and D-3PA traces are given, which corresponds to the presence of only the signal or the pump. Indeed, the residual D-3PA background was negligible. In contrast to the imaging based on the raster scanning \cite{Pattanaik2016OE, Wang2012OE}, the direct two-dimensional imaging was featured with improved simplicity and speed. As presented in Fig. \ref{fig2}(c), the count rate by summing up all the registered photons in the image exhibited expected linear and quadratic dependences on the incident signal power for the non-degenerated (ND-2PA) and degenerated (D-2PA) regimes, respectively. Furthermore, the enhancement factor could be calculated, which was inversely proportional to the signal pulse energy as given in Fig. \ref{fig2}(d). For the aforementioned power settings, the enhancement factor was about 32. In principle, the enhancement would be more pronounced with a weaker signal power, albeit with a relatively larger uncertainty due to the reduced photon counts.

\begin{figure}[t!]
\centering
\includegraphics[width=0.95\columnwidth]{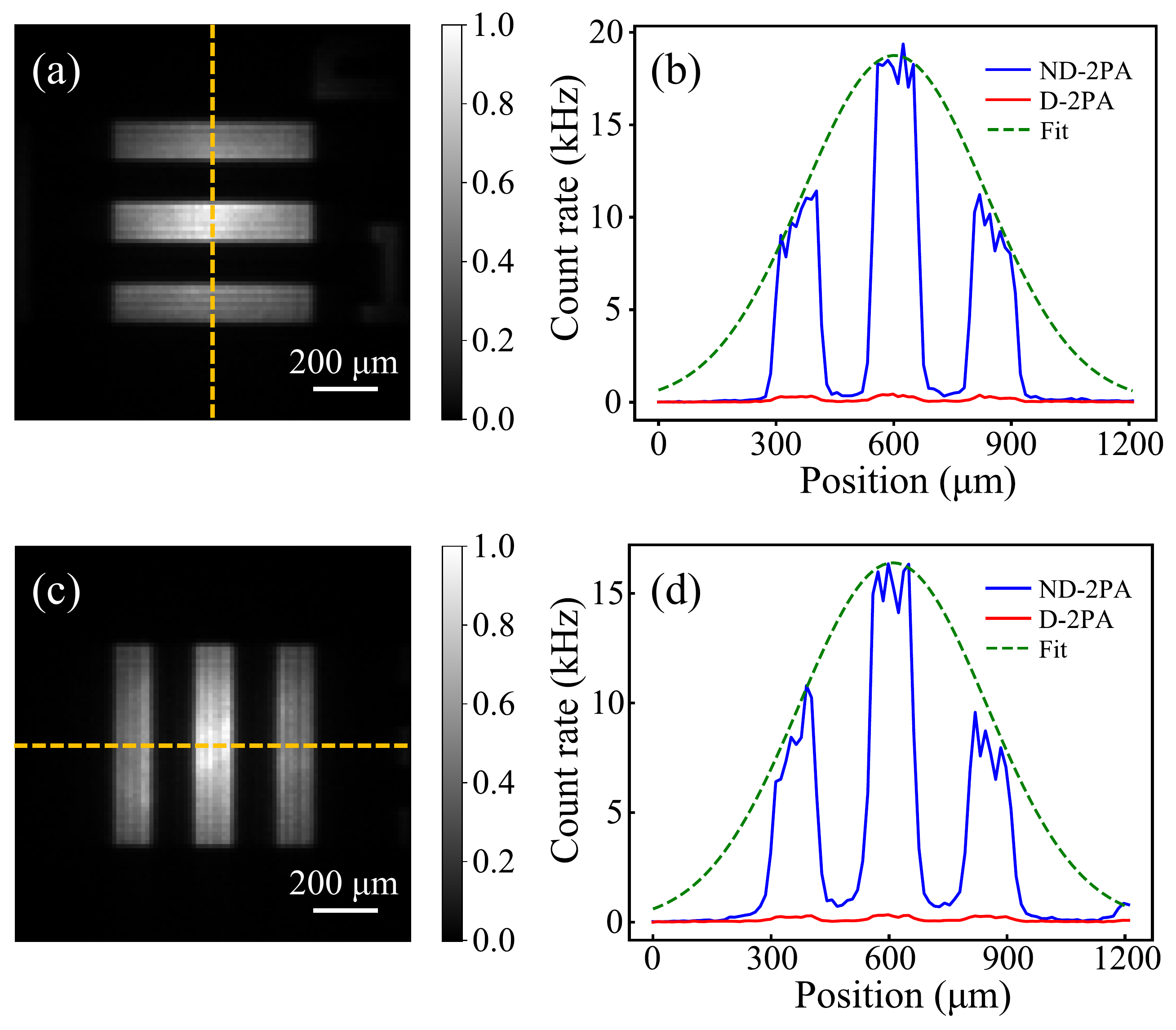}
\caption{Recorded images with ND-2PA for the horizontal (a) and vertical (c) three-bar patterns for the first element of group 2 in the USAF-1951 resolution target. The photon-counting distributions along the dashed lines in the image are presented in (b) and (d). The relevant intensity distributions for the recorded images with the D-2PA are also given as the direct comparison.}
\label{fig3}
\end{figure}

\begin{figure}[t!]
\centering
\includegraphics[width=0.7\columnwidth]{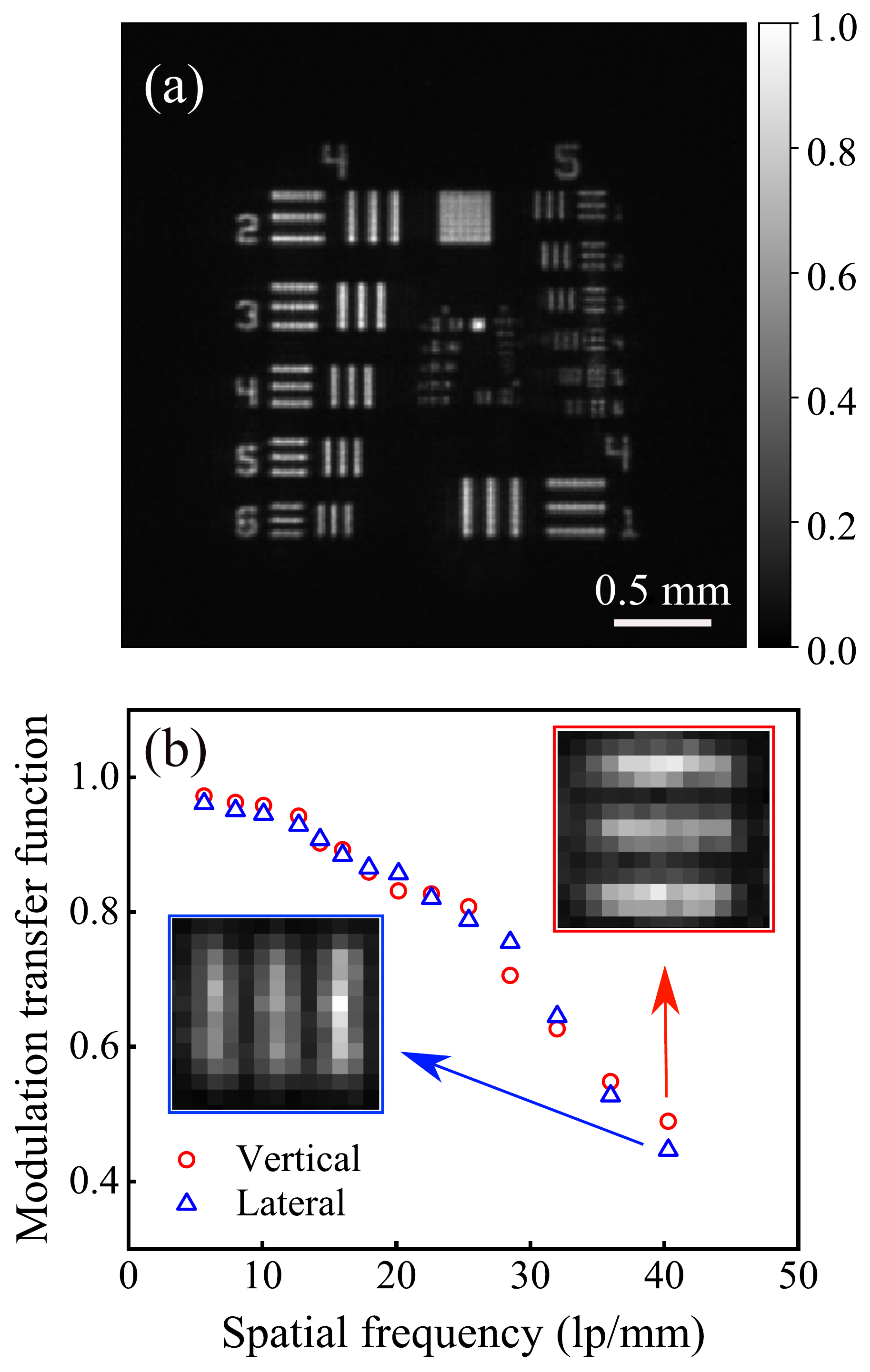}
\caption{(a) Stitched image for 16 acquisitions with the spatially scanned pump beam. The resulting field of view is enlarged by three times in this proof-of-principle demonstration. (b) Imaging contrast as a function of the spatial frequency. Two recorded images in the inset correspond to 40.3 lines per mm, indicating a spatial resolution of 12.4 $\mu$m.}
\label{fig4}
\end{figure}

To investigate the imaging resolution, a USAF-1951 resolution target (Thorlabs, R3L3S1N) was inserted in the signal path. The resultant object image and the CCD camera constituted a 4f imaging system by using two lenses. Figures \ref{fig3}(a) and (c) present the recorded images for the horizontal and vertical three lines (Group 2, Element 1) in the ND-2PA scenario. The corresponding intensity distributions are given in Figs. \ref{fig3}(b) and (d). The reduced heights for two side bars are due to the limited beam size of the signal beam. Meanwhile, the image area on the CCD is determined by the pump size. In general, a larger pump beam would broaden the detected field of view, but the reduced light intensity would in turn limit the detection sensitivity. Due to the practical trade-off, the diameter of the pump beam was engineered to be about 1.4 mm in the experiment. It is worth noting that for the 2PA-based imaging system, the field of view and the spatial resolution are defined by the imaging system itself rather than the pump beam. For a given pump beam size, the detected field of view can be enlarged by scanning the pump positions on the CCD. As a proof-of-principle demonstration, Fig. \ref{fig4}(a) presents the stitched image for 4 $\times$ 4 acquisitions with the raster-scanned pump beam. Consequently, the field of view could be effectively increased by three times. In combination with high-speed scanning galvanometers, a wide-angle imaging at a video frame rate would be possible. 

To facilitate the quantitative analysis on the imaging quality, the so-called modulation transfer function (MTF) was used, which was related to the image contrast or visibility. For the imaged pattern with 4 line pairs per millimeter (lp/mm) in Figs. \ref{fig3}(a,c), the MTF was calculated to be 0.98 and 0.96 for the vertical and lateral resolving abilities. As indicated in Fig. \ref{fig4}(b), the spatial frequency for the imaging system could approach to about 40 lp/mm, which corresponded to a resolution closed to the pixel size of 13 $\mu$m. Notably, the spatial resolution could be substantially improved to the diffraction limit by using proper lenses with a large aperture, as been routinely achieved in a common microscope. Such a superior resolving capability benefited from the on-chip nonlinearity for the two-photon absorption, which was beyond the phase-matching restricted performance in the upconversion imaging based on nonlinear crystals \cite{Huang2012APL, Barh2019AOP}.

\begin{figure}[t!]
\centering
\includegraphics[width=0.9\columnwidth]{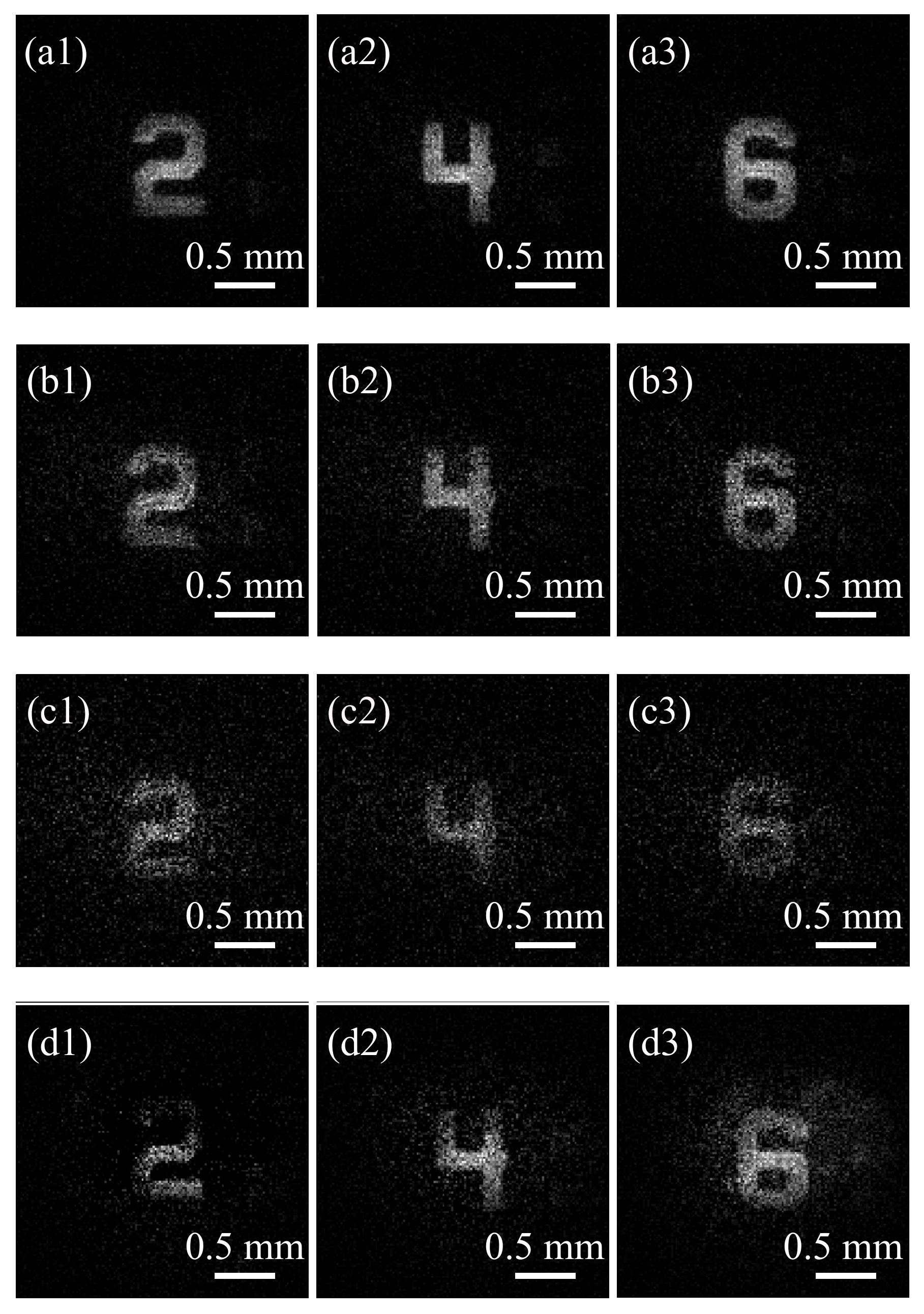}
\caption{Captured images based on the ND-2PA for various incident pulse energies. The images in (a1-a3), (b1-b3) and (c1-c3) correspond to the incident illumination of 100, 50 and 25 photons/pixel/pulse, respectively. The exposure time is set to be 8 s. (d1-d3) present the single-photon operation with 1 photon/pixel/pulse albeit with a longer accumulation time of 80 s.}
\label{fig5}
\end{figure}

In the next, we investigated the sensitivity of the imaging system. To this end, the EMCCD was cooled down to -80 $^\circ$C by using the build-in thermoelectric cooling, and operated at the 1000-fold high gain and 30-MHz readout rate. The dark current of the EMCCD was specified to be about 10$^{-4}$ electrons/pixel/s, which permitted a highly sensitive imaging at the single-photon level. Figure \ref{fig5} presents the captured images based on the ND-2PA for various impinging pulse energies. Specifically, Figs. \ref{fig5}(a1-a3), (b1-b3) and (c1-c3) correspond to the incident illumination of 100, 50 and 25 photons/pixel/pulse, respectively. The exposure time was set to be 8 s. As shown in Figs. \ref{fig5}(d1-d3), further reduction of the photon number to 1 was still allowed to obtain high-contrast images with a longer accumulation time up to 80 s. The achieved single-photon operation thus represented the benchmark in the imaging sensitivity based on the 2PA scheme.

\begin{figure}[t!]
\centering
\includegraphics[width=0.95\columnwidth]{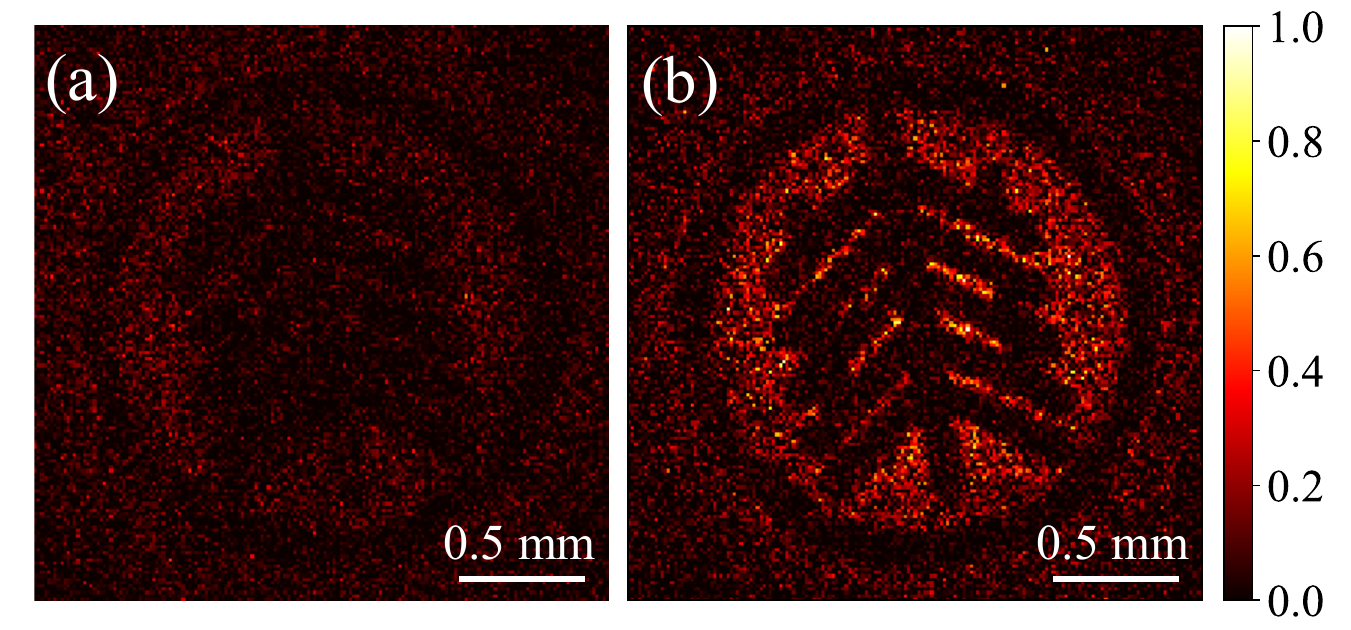}
\caption{Infrared imaging of a silicon wafer engraved with the university logo under the same illumination power. (a) and (b) correspond to the D-2PA and ND-2PA scenarios.}
\label{fig6}
\end{figure}

\begin{figure*}[t!]
\centering
\includegraphics[width=0.75\textwidth]{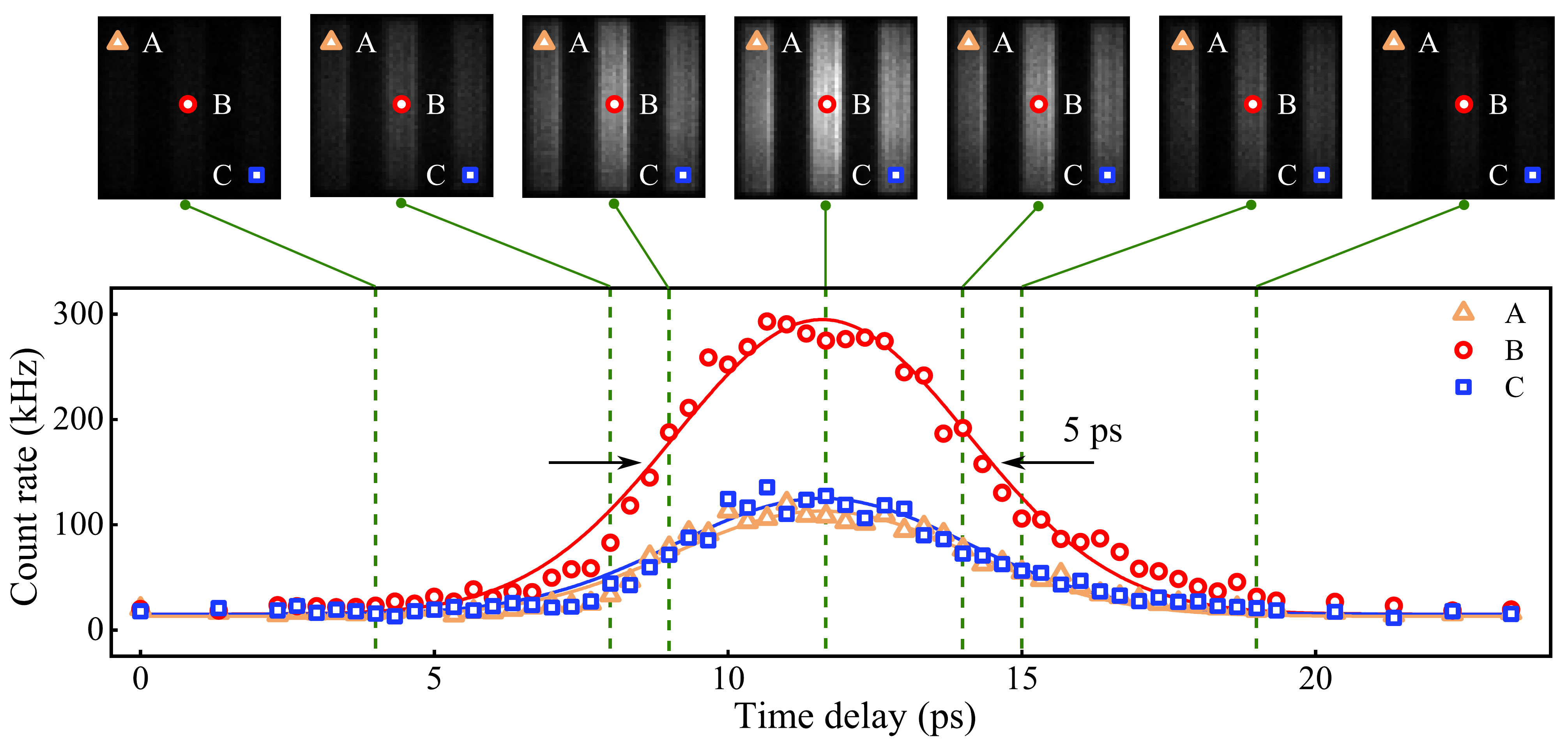}
\caption{Detected count rates for three selected pixels vary as a function of the temporal delay between the pump and signal pulses. The images on the top correspond to typical intensity distributions at difference time delay.}
\label{fig7}
\end{figure*}

Finally, we demonstrated the imaging capability through the semiconductor samples. The resolution test target was replaced by a silicon wafer. The wafer had a thickness of 200 $\mu$m and was polished at the two surfaces. Particularly, the backside was inscribed with the university logo by using the laser ablation. There would be stronger scattering at the ablated part, thus leading to a reduced power transmission. The transmitted infrared light could generate the object image, which was then mapped onto the CCD camera. Figures \ref{fig6} (a) and (b) correspond to the D-2PA and ND-2PA scenarios, which shows the enhanced signal-to-noise ratio in the presence of MIR pumping. The sensitive infrared imaging at the telecom wavelength might find application in non-destructive defect inspection for the semiconductor chips \cite{Pattanaik2016OE}. Additionally, the temporal resolution of the imaging system was investigated by tuning the delay line (Delay1 shown in Fig. \ref{fig1}). As shown on the top of Fig. \ref{fig7}, the images correspond to intensity distributions at a given time delay. The detected count rates for three selected pixels (denoted by A, B, and C) varied as a function of the temporal delay between the pump and signal pulses. The resulting cross-correlation trace exhibited a time width about 5 ps, which would allow a longitudinal resolution of 0.75 mm for a reflective imaging. The depth resolution is possible to reach the $\mu$m level by using femtosecond ultrafast pulses. It is worth noting that the optical gating technique would permit a temporal resolution orders of magnitude lower than the typical timing jitter by direct detection \cite{Zhang2015LSA}.

\section{Conclusion}
In conclusion, we have experimentally implemented an ultra-sensitive infrared imaging system based on the ND-2PA. The adopted long-wavelength-pumping technique facilitated the non-degenerated operation, which would improve the nonlinear absorption coefficient for the signal. In the meanwhile, the elimination of second-harmonic noises of the intensive pump enabled to significantly increase the signal-to-noise ratio. Consequently, the single-photon sensitivity was unprecedentedly achieved, thus representing the benchmark in the imaging sensitivity based on the 2PA scheme. To go beyond the achieved performance, a longer pump wavelength over 3300 nm can be employed, which could further exclude the residual 3PA noises \cite{Hayat2008PRB}. In this scenario, higher-power ultrafast pump with a larger beam size would be preferable to approach the infrared imaging with high sensitivity, large field of view, as well as high resolution in the transverse and longitudinal dimensions. The desirable imaging capability would stimulate applications including time-resolved microscopy, time-of-flight 3D imaging, and volumetric optical tomography \cite{Bruschini2019LSA, Pavia2014JSTQ}.

\begin{IEEEbiographynophoto}{Jianan Fang}
was born in Zhejiang Province, China, in 1997. He received the B.S. degree from Hangzhou Normal University, Zhejiang, China, in 2019, where he is currently working toward the Ph.D. degree. His research interest is focused on the infrared detection and imaging.
\end{IEEEbiographynophoto}

\begin{IEEEbiographynophoto}{Yinqi Wang}
was born in Liaoning Province, China, in 1996. She received the B.S. degree from University of Shanghai for Science and Technology, Shanghai, China, in 2018, where she is currently working toward the Ph.D. degree. Her research interest is focused on the optical nonlinear frequency conversion.
\end{IEEEbiographynophoto}

\begin{IEEEbiographynophoto}{E Wu}
was born in Shandong Province, China, in 1979. She received her B.S. in 2001 from Department of Physics at East China Normal University and Ph.D. degree in 2007 from both East China Normal University and \'{E}cole Normale Sup\'{e}rieure de Cachan (France) as a cotutored student. After graduation, she joined State Key Laboratory of Precision Spectroscopy, East China Normal University as Associate Professor, where she became Researcher at the beginning of 2014. She focuses her research in single-photon generation, frequency conversion and interference.
\end{IEEEbiographynophoto}

\begin{IEEEbiographynophoto}{Ming Yan}
was born in Shaanxi Province, China, in 1984. He received a B.S. in Electronics Engineering (2007) and a Ph. D degree in Optics (2012) at East China Normal University (Shanghai, China). He was a postdoc researcher at Max-Planck Institute of Quantum Optics (Garching Germany, 2013-2017), and then joined the State Key Laboratory of Precision Spectroscopy, East China Normal University in 2017 as an associate professor. In 2019, he was promoted to be a professor. His research interests include fiber lasers, optical frequency combs and molecular spectroscopy.
\end{IEEEbiographynophoto}

\begin{IEEEbiographynophoto}{Kun Huang}
was born in Jiangxi Province, China, in 1986. He received PhDs from \'{E}cole Normale Sup\'{e}rieure de Paris (France) and East China Normal University in 2015. Then he did postdoc in Pierre-and-Marie-Curie University before joining University of Shanghai for Science and Technology as a Professor in 2017. Since 2019, he has been a Researcher in State Key Laboratory of Precision Spectroscopy, East China Normal University.  He has coauthored more than 50 articles in peer-reviewed journals and 3 Chinese patents. His current research interests include single-photon detection, precision spectroscopy, fiber lasers, and quantum optics.
\end{IEEEbiographynophoto}

\begin{IEEEbiographynophoto}{Heping Zeng}
was born in Hunan Province, China, in 1966. He received the B.S. degree in physics from Peking University, Beijing, China, in 1990. And then he obtained the Ph.D. degree from the Shanghai Institute of Optics and Fine Mechanics, Chinese Academy of Science, Shanghai, China, in 1995. In 2000, he joined East China Normal University as a Professor. In 2016, he was selected as OSA Fellow. His current research interests were focused on precision spectroscopy and quantum detection. He has published more than 300 academic papers in peer-reviewed journals. 
\end{IEEEbiographynophoto}

\end{document}